\documentclass[prd]{revtex4}
\usepackage{amsmath}
\usepackage{amssymb}
\usepackage{graphics}
\usepackage{epsfig}

\newcommand\ba{\begin{eqnarray}}
\newcommand\ea{\end{eqnarray}}
\newcommand\nn{\nonumber}

\newcommand{\br}[1]{\left( #1 \right)}
\newcommand{\brs}[1]{\left[ #1 \right]}

\newcommand{\brm}[1]{\left| #1 \right|}
\newcommand{\GeV}{~\mbox{GeV}}
\newcommand{\MeV}{~\mbox{MeV}}
\newcommand{\KeV}{~\mbox{KeV}}

\begin{document}

\title{$2\gamma$-decays of scalar mesons ($\sigma(600)$, $f_0(980)$ and $a_0(980)$)
in the Nambu-Jona-Lasinio model}

\author{M.~K.~Volkov}
\email{volkov@theor.jinr.ru}
\affiliation{Joint Institute for Nuclear Research, Dubna, Russia}

\author{Yu.~M.~Bystritskiy}
\email{bystr@theor.jinr.ru}
\affiliation{Joint Institute for Nuclear Research, Dubna, Russia}

\author{E.~A.~Kuraev}
\email{kuraev@theor.jinr.ru}
\affiliation{Joint Institute for Nuclear Research, Dubna, Russia}

\begin{abstract}
The two-photon decay widths of scalar mesons
$\sigma(600)$, $f_0(980)$ and $a_0(980)$ are calculated in framework of
the local Nambu-Jona-Lasinio model. The contributions
of the quark loops (Hartree-Fock approximation) and the meson
loops (next $1/N_c$-approximation where $N_c$ is the number of colors)
are taken into account.
These contributions, as we show, are the values of the same order of magnitude.
For the $f_0$ decay the $K$-loop contribution turns out to play the
dominant role.
The results are in satisfactory agreement with modern experimental
data.
\end{abstract}

\maketitle

\section{Introduction}

In recent papers \cite{Bystritskiy:2007wq,Volkov:2008ye}, the radiative decay
$\phi\to f_0\gamma$ and $f_0 (a_0) \to \rho(\omega) \gamma$ widths
within the local Nambu-Jona-Lasinio (NJL) model
\cite{Volkov:1986zb,Volkov:1982zx,Ebert:1982pk,Volkov:1984kq,Ebert:1985kz,Ebert:1994mf,Volkov:2006vq,Vogl:1991qt,Klevansky:1992qe}
have been calculated.
In these works, we took into account not only
the quark loop contributions but also the meson
loop contributions, moreover,
in the decays of the $f_0(980)$ meson the kaon loop contribution is the
dominant one.
It is worth noticing that the situation here is
similar to the one that takes place in the case of
$\phi\to f_0 \gamma$, $f_0\to\rho\gamma$, $f_0\to\omega\gamma$ decays.
Let us note that due to the explicit gauge invariant form of the amplitude
the relevant loop integrals do not contain ultraviolet divergences.
Thus, the explicit dependence of these amplitudes of the external momenta
was obtained \cite{Ebert:1996pc}.

In this paper, we will consider the two-photon decays of the scalar mesons
$\sigma(600)$, $f_0(980)$ and $a_0(980)$. These decays was
considered in a recent paper \cite{Kalinovsky:2008iz} where
a rather rough $q^2$ approximation for quark loop integrals evaluation was used.
In our case we consider the full integral corresponding to the
quark loop and thus, we can obtain a complete dependence of the amplitude of
external momenta.

In the case of the quark loop we consider only a real part of the
relevant loop integral.
This prescription permits us to take into account the condition of
the "naive" quark confinement. Some theoretical arguments
supporting this procedure can be found in \cite{Pervushin:1985yi}.
As for the meson loops, both the real and the imaginary parts were
taken into account.

The structure of our paper is the following.
In Section~\ref{Lagrangian}, the NJL quark-meson Lagrangian,
corresponding parameters and the coupling constants of our model
are defined.
In Section~\ref{LoopIntegrals}, the methods of
quark and meson loop calculation are given.

In Section~\ref{Amplitudes}, the contributions of quark and meson loops
to the amplitudes and the widths of
two-photon decays of the scalar meson are presented.

In Section~\ref{Conclusion}, we discuss the results obtaned.

\section{Lagrangian of the NJL model}
\label{Lagrangian}

The Lagrangian of interaction of mesons and quarks within the
NJL model has the form \cite{Volkov:1986zb}:
\ba
    {\cal L} &=&
    \bar q \left[ i \hat\partial-M+e Q \hat A +
    g_{\sigma_u} \lambda_u \sigma_u + g_{\sigma_s} \lambda_s \sigma_s + g_u \lambda_3 a_0
    +\right. \nn\\
    &&\qquad+\left.
        i \gamma_5 g_\pi \br{\lambda_{\pi^+} \pi^+ + \lambda_{\pi^-} \pi^-}
        +
        i \gamma_5 g_K \br{\lambda_{K^+} K^+ + \lambda_{K^-} K^-}
        \frac{}{}
    \right] q,
    \label{QuarkMesonLagrangian}
\ea
where $\bar q= \br{\bar u,\bar d,\bar s}$, and $u$, $d$, $s$ are the quark fields,
$M=\mbox{diag}\br{m_u,m_d,m_s}$ with $m_u=m_d=263\MeV$, $m_s=406\MeV$ -
constituent quark mass matrix;
$Q=\mbox{diag}\br{2/3,-1/3,-1/3}$ is the quark electric charge matrix,
$e$ is the elementary electric charge ($e^2/4\pi=\alpha=1/137$),
$\lambda_u=\br{\sqrt{2} \lambda_0+\lambda_8}/\sqrt{3}$,
$\lambda_s=\br{-\lambda_0+\sqrt{2}\lambda_8}/\sqrt{3}$,
$\lambda_{\pi^\pm}=(\lambda_1\pm i\lambda_2)/\sqrt{2}$,
$\lambda_{K^\pm}=(\lambda_4\pm i\lambda_5)/\sqrt{2}$
where $\lambda_i$ are the well-known Gell-Mann matrices and
$\lambda_0 = \sqrt{2/3}~\mbox{diag}\br{1,1,1}$.
Scalar isoscalar mesons $f_0$, $\sigma$ are the mixed states
\ba
f_0 &=& \sigma_u\sin\alpha+\sigma_s\cos\alpha, \nn \\
\sigma &=& \sigma_u\cos\alpha-\sigma_s\sin\alpha,
\ea
with the mixing angle $\alpha=11.3^o$ \cite{Volkov:1999qb,Volkov:1999qn,Volkov:2006vq}.

The coupling constants from the Lagrangian (\ref{QuarkMesonLagrangian}) are
defined in the following way \cite{Volkov:1986zb}:
\ba
g_{\sigma_u} &=& \left( 4 I^\Lambda\br{m_u, m_u}\right)^{-1/2} = 2.43, \nn\\
g_{\sigma_s} &=& \left( 4 I^\Lambda\br{m_s, m_s}\right)^{-1/2} = 2.99, \nn \\
g_{\pi} &=& \frac{m_u}{F_\pi} = 2.84, \nn\\
g_K &=& \frac{m_u+m_s}{2 F_K} = 3.01, \nn
\ea
where we use the Goldberger-Treiman relation for $g_{\pi}$ and $g_K$ constants,
$F_\pi=92.5\MeV$ and $F_K = 1.2~F_\pi$, and $I^\Lambda\br{m, m}$ is the
logarithmically divergent integral which has the form:
\ba
    I(m,m) &=& \frac{N_c}{\br{2\pi}^4}
    \int d^4 k
    \frac{\theta\br{\Lambda^2-k^2}}
    {\br{k^2+m^2}^2} =
    \frac{N_c}{\br{4\pi}^2}
    \br{
        \ln\br{\frac{\Lambda^2}{m^2}+1}
        -
        \frac{\Lambda^2}{\Lambda^2 + m^2}
    }, \qquad N_c = 3. \nn
\ea
This integral is written in the Euclidean space.
The cut-off parameter $\Lambda = 1.27\GeV$
is taken from \cite{Volkov:1986zb,Bystritskiy:2007wq}.

\section{Quark and meson loop integrals}
\label{LoopIntegrals}

The amplitudes of the $2\gamma$ decay can be expressed in terms of
the quark and meson loop integrals.

The quark loop contribution to the amplitude is given by
two triangle type Feynman diagrams:
\ba
    T_{\mu\nu}^q &=&
    -\frac{\alpha}{4\pi}
    \int \frac{d^4k}{i \pi^2}
    \frac{
        Sp\brs{
            \gamma_\nu\br{\hat k + m_q}\gamma_\mu\br{\hat k + \hat q_1 + m_q}\br{\hat k - \hat q_2 + m_q}
        }
    }
    {
        \br{k^2-m_q^2}
        \br{\br{k+q_1}^2-m_q^2}
        \br{\br{k-q_2}^2-m_q^2}
    }
    +\br{\br{q_1,\mu} \leftrightarrow \br{q_2,\nu}}.
    \ea
Applying the Feynman procedure of joining of the denominators
\ba
\frac{1}
    {
        \br{k^2-m_q^2}
        \br{\br{k+q_1}^2-m_q^2}
        \br{\br{k-q_2}^2-m_q^2}
    }
    =
    \int\limits_0^1 dx
    \int\limits_0^1 2 y dy
    \frac{1}{\br{\br{k-q_x y}^2 - \br{m_q^2 + q_x^2 y^2}}^3},
\ea
where $q_x = x q_2 - \bar xq_1$, $\bar x = 1-x$, we obtain for
$T_{\mu\nu}^q$:
\ba
    T_{\mu\nu}^q &=&
        \frac{\alpha}{\pi}
        \br{g_{\mu\nu}\br{q_1 q_2} - q_{1\nu} q_{2\mu}} T^q, \\
    T^{q} &=& 2m_q
    \int\limits_0^1 dx
    \int\limits_0^1 y dy
    \frac{1-4y^2 x \bar x}{m_q^2 - M_S^2 y^2 x \bar x}.
\ea

For meson loops an additional Feynman diagram with two photon-two meson vertex
contributes as well. To restore the general gauge invariant form of the amplitude,
we can nevertheless consider only two triangle type Feynman diagrams:
\ba
    \Delta T_{\mu\nu}^M &=&
    \frac{\alpha}{4\pi}
    \int \frac{d^4k}{i \pi^2}
    \frac{
        \br{2k+q_1}_\mu \br{2k-q_2}_\nu
    }
    {
        \br{k^2-M^2}
        \br{\br{k+q_1}^2-M^2}
        \br{\br{k-q_2}^2-M^2}
    }
    +\br{\br{q_1,\mu} \leftrightarrow \br{q_2,\nu}}.
\ea
Extracting the term $\sim q_{1\nu}q_{2\mu}$ and adding the relevant term $\sim g_{\mu\nu}$
we obtain:

\ba
    T_{\mu\nu}^M &=&
        \frac{\alpha}{\pi}
        \br{g_{\mu\nu}\br{q_1 q_2} - q_{1\nu} q_{2\mu}} T^M, \\
\ea
with
\ba
 T^{M} &=& 2
    \int\limits_0^1 dx
    \int\limits_0^1 y dy
    \frac{y^2 x \bar x}{M^2 - M_S^2 y^2 x \bar x}.
\ea
Standard evaluation of these integrals leads to
\ba
    T^{q} &=& -\frac{1}{m_q} F(z^q_S), \\
    T^{M} &=& -\frac{z^M_S}{4M^2} \Phi(z^M_S),
\ea
where $z^q_S = 4 m_q^2/M_S^2$, $z^M_S = 4 M^2/M_S^2$,
\ba
    F(z) &=& \mbox{Re}\brs{1+\br{1-z}\Phi(z)}, \nn\\
    \Phi(z) &=& z \phi(z) - 1, \nn\\
    \phi(z) &=&
    \left\{
        \begin{array}{ll}
            \frac{1}{4} \brs{\pi^2 - \ln^2 \frac{1+\sqrt{1-z}}{1-\sqrt{1-z}}} +
            i \frac{\pi}{2} \ln \frac{1+\sqrt{1-z}}{1-\sqrt{1-z}}, & z < 1, \\
            \br{\arctan \frac{1}{\sqrt{z-1}}}^2, & z>1. \\
        \end{array}
    \right.
\ea
We remind that for the quark loop contribution
the imaginary part of the function $\Phi(z)$ must be omitted,
and for the meson loop contribution both the real and
possible imaginary parts are relevant.

Similar expressions were obtained in \cite{Ebert:1996pc},
where the imaginary part of the quark loop contribution was
taken into account.

\section{Scalar mesons decay amplitudes and widths}
\label{Amplitudes}

The vertices of the quark-meson and quark-photon interactions were given above.
The vertices of the meson-meson interaction in the framework of
NJL model have the form (for details see \cite{Volkov:1986zb}):
\ba
    V_{\sigma_s \pi^+\pi^-} &=& V_{a_0 \pi^+\pi^-} = 0, \nn\\
    V_{\sigma_u K^+K^-} &=& V_{a_0 K^+K^-} = -2\br{2m_u - m_s} \frac{g_K^2}{g_{\sigma_u}}, \nn\\
    V_{\sigma_s K^+K^-} &=& 2\sqrt{2}\br{2m_s - m_u} \frac{g_K^2}{g_{\sigma_s}}, \nn\\
    V_{\sigma_u \pi^+\pi^-} &=& -4m_u \frac{g_\pi^2}{g_{\sigma_u}}. \nn
\ea
The general structure of the two-photon scalar meson decay amplitudes has the form
\ba
T_{S\gamma\gamma}=-\frac{\alpha g_{\sigma_u}}{\pi m_u}\br{g_{\mu\nu}\br{q_1 q_2} - q_1^\nu q_2^\mu}a_{S\gamma\gamma}.
\ea
The expression for the width has the form:
\ba
\Gamma_{S\gamma\gamma}=\frac{M_S^3}{64\pi}\frac{\alpha^2g^2_{\sigma_u}}{\pi^2m_u^2}|a_{S\gamma\gamma}|^2.
\ea

The amplitude $a_{a_0\gamma\gamma}$ of $a_0\to \gamma\gamma$ contains the contribution of
$u$, $d$ quarks and the $K$-meson intermediate states.
The color-charge factor associated with $u$, $d$ quarks is $N_c\br{\frac{4}{9}-\frac{1}{9}}=1$.
Thus,
\ba
a^{u,d}_{a_0\gamma\gamma} &=&
 F\br{z_{a_0}^u}.
\ea
Taking the $K$-meson loop contribution we obtain:
\ba
a_{a_0\gamma\gamma} &=&
    a^{u,d}_{a_0\gamma\gamma} + a^{K}_{a_0\gamma\gamma} = \nn\\
    &=&
    F\br{z_{a_0}^u}
    -
    \frac{m_u}{g_{\sigma_u}}
    \frac{2\br{2m_u-m_s} g_K^2}{4 g_{\sigma_u} M_K^2}
    z_{a_0}^K \Phi\br{z_{a_0}^K}=0.482-0.114=0.367.
\ea
The corresponding width is
\ba
\Gamma_{a_0(980)\to\gamma\gamma} &=& \br{2.25\KeV} \brm{a_{a_0\gamma\gamma}}^2 = 0.29\KeV. \nn
\ea

In the case of the $f_0\to \gamma\gamma$ decay we also have the contribution of
$u$, $d$ and $s$ quarks and the $K$-meson intermediate states.
The color-charge factor associated with $u$, $d$ quarks is $N_c\br{\frac{4}{9}+\frac{1}{9}}=\frac{5}{3}$
for $\sigma_u$-component of $f_0$ and $N_c\frac{1}{9}=\frac{1}{3}$
for $\sigma_s$-component of $f_0$.
Taking the $K$-meson and the $\pi$-meson loop contribution we obtain
\ba
a_{f_0\to \gamma\gamma} &=&
\frac{5}{3}F\br{z_{f_0}^u} \sin\alpha
-
\frac{\sqrt{2}}{3}F\br{z_{f_0}^s}\br{\frac{g_{\sigma_s} m_u}{g_{\sigma_u} m_s}} \cos\alpha
+
\nn\\
&+&
\br{
    -\frac{g_K^2}{g_{\sigma_u}^2}
    \frac{m_u}{4 M_K^2}
    2\br{2 m_u-m_s}\sin\alpha
    +
    \frac{g_K^2}{g_{\sigma_u}g_{\sigma_s}}
    \frac{m_u}{4 M_K^2}
    2\sqrt{2}\br{2 m_s-m_u}\cos\alpha
}
z_{f_0}^K \Phi\br{z_{f_0}^K}
- \nn\\
&-&
    \sin\alpha
    \frac{m_u^2}{M_\pi^2}
    \frac{g_\pi^2}{g_{\sigma_u}^2}
    z_{f_0}^\pi \Phi\br{z_{f_0}^\pi}=0.157-0.417-0.022+0.589+0.082-0.038 i=0.385-0.038 i.
\ea
For the width we have
\ba
\Gamma_{f_0(980)\to\gamma\gamma} &=& \br{2.25\KeV} \brm{a_{f_0\gamma\gamma}}^2 = 0.33\KeV. \nn
\ea

In the case of the $\sigma\to \gamma\gamma$ decay we have
\ba
a_{\sigma\to \gamma\gamma} &=&
\frac{5}{3}F\br{z_{\sigma}^u} \cos\alpha
+
\frac{\sqrt{2}}{3}F\br{z_{\sigma}^s}\br{\frac{g_{\sigma_s} m_u}{g_{\sigma_u} m_s}} \sin\alpha
\nn\\
&-&
\br{
    \frac{g_K^2}{g_{\sigma_u}^2}
    \frac{m_u}{4 M_K^2}
    2\br{2 m_u-m_s}\cos\alpha
    +
    \frac{g_K^2}{g_{\sigma_u}g_{\sigma_s}}
    \frac{m_u}{4 M_K^2}
    2\sqrt{2}\br{2 m_s-m_u}\sin\alpha
}
z_{\sigma}^K \Phi\br{z_{\sigma}^K}
-\nn\\
&-&
    \cos\alpha
    \frac{m_u^2}{M_\pi^2}
    \frac{g_\pi^2}{g_{\sigma_u}^2}
    z_{\sigma}^\pi \Phi\br{z_{\sigma}^\pi}=1.89+0.057-0.041-0.043+0.92-0.98 i=2.78 - 0.98 i
\ea

The corresponding width is
\ba
\Gamma_{\sigma(600)\to\gamma\gamma} &=& \br{0.51\KeV} \brm{a_{\sigma\gamma\gamma}}^2 = 4.3\KeV.
\ea

The experimental value of the mass and the width of the $\sigma$ meson is not well established.
We present the width of $\sigma$ for two other masses: $M_\sigma=450\MeV$ and $M_\sigma=550\MeV$.
They are
\ba
\Gamma_{\sigma(450)\to\gamma\gamma}= 2.18\KeV,\nn \\
\Gamma_{\sigma(550)\to\gamma\gamma}= 3.53\KeV.\nn
\ea
The comparison of our results with the experimental data and some other model
predictions is given in Table~\ref{TableOfDecays}.

\begin{table}
\begin{tabular}{|l|l|}
\hline
\hline
$\Gamma\br{a_0\to\gamma\gamma}\br{exp.}$, $\KeV$ &
$\Gamma\br{a_0\to\gamma\gamma}\br{theor.}$, $\KeV$ \\
\hline
\hline
$0.30\pm 0.10$ \cite{Amsler:1997up} & $0.29$ (This paper)\\
\hline
\hline
$\Gamma\br{f_0\to\gamma\gamma}\br{exp.}$, $\KeV$ &
$\Gamma\br{f_0\to\gamma\gamma}\br{theor.}$, $\KeV$ \\
\hline
\hline
$0.205^{+0.095+0.147}_{-0.083-0.117}$ \cite{Mori:2006jj} & $0.33$ (This paper) \\
\hline
$0.28^{+0.09}_{-0.13}$ \cite{Boglione:1998rw} & $0.21-0.26$ \cite{Branz:2008ha} \\
\hline
$0.42 \pm 0.06 \pm 0.18$ \cite{Oest:1990ki} & $0.22$ \cite{Hanhart:2007wa} \\
\hline
$0.29 \pm 0.07 \pm 0.12$ \cite{Boyer:1990vu} & $0.33$ \cite{Schumacher:2006cy} \\
\hline
$0.31 \pm 0.14 \pm 0.09$ \cite{Marsiske:1990hx} & $0.31$ \cite{Scadron:2003yg} \\
\hline
$0.63 \pm 0.14$ \cite{Morgan:1990kw} & $0.28^{+0.09}_{-0.13}$ \cite{Anisovich:2001zp} \\
\hline
                                     & $0.20$ \cite{Oller:1997yg} \\
\hline
                                     & $0.24$ \cite{Efimov:1993zg} \\
\hline
                                     & $0.27$ \cite{Achasov:1981kh} \\
\hline
\hline
$\Gamma\br{\sigma\to\gamma\gamma}\br{exp.}$, $\KeV$ &
$\Gamma\br{\sigma\to\gamma\gamma}\br{theor.}$, $\KeV$ \\
\hline
\hline
$4.1 \pm 0.3$ \cite{Pennington:2006dg} & $4.3$ (This paper)\\
\hline
$3.8 \pm 1.5$ \cite{Boglione:1998rw} & \\
\hline
$5.4 \pm 2.3$ \cite{Morgan:1990kw} & \\
\hline
$10 \pm 6$ \cite{Courau:1986gn} & \\
\hline
\end{tabular}
\caption{The table of two-gamma decays of the scalar mesons
$\sigma(600)$, $f_0(980)$ and $a_0(980)$.
\label{TableOfDecays}}
\end{table}

\section{Conclusion}
\label{Conclusion}

The calculations of the radiative decays in the NJL model show an important
role of both the quarks
and meson loops. Moreover, for the $f_0$ meson decay the kaon loop turns out to
provide the dominant contribution.
It is worth noticing that the situation here is
similar to the one that takes place in the case of
$\phi\to f_0 \gamma$ \cite{Bystritskiy:2007wq}, $f_0\to\rho(\omega)\gamma$ \cite{Volkov:2008ye}
decays.
This fact permits one to understand the success
of such models as the model of a kaon molecule \cite{Weinstein:1990gu} as well as the
four-quark model \cite{Achasov:1987ts,Achasov:2008ut,Branz:2008ha}.
The NJL model used here allows us to take into account
both the quark-antiquark state, which manifests itself in the form of quark loops,
and the hidden four-quark state, which shows up as meson loops.
Let us emphasize that in the framework of the standard NJL model we can describe the
$2\gamma$ decays without any additional parameters.

\begin{acknowledgements}
The authors wish to thank Prof. N. N. Achasov, Prof. S. B. Gerasimov
and Prof. V. N. Pervushin
for fruitful discussions.
We also acknowledge the support of INTAS grant no. 05-1000008-8528.
\end{acknowledgements}


\end{document}